\documentclass[conference]{IEEEtran}
\IEEEoverridecommandlockouts
\usepackage{cite}
\usepackage{amsmath,amssymb,amsfonts}
\usepackage{algorithmic}
\usepackage[ruled,linesnumbered]{algorithm2e}
\usepackage{graphicx}
\ifCLASSOPTIONcompsoc
    \usepackage[caption=false, font=normalsize, labelfont=sf, textfont=sf]{subfig}
\else
\usepackage[caption=false, font=footnotesize]{subfig}
\fi
\usepackage{textcomp}
\usepackage{xcolor}
\def\BibTeX{{\rm B\kern-.05em{\sc i\kern-.025em b}\kern-.08em
    T\kern-.1667em\lower.7ex\hbox{E}\kern-.125emX}}
\begin{document}

\title{Deep Unrolled Network for Video Super-Resolution\\

}

\author{\IEEEauthorblockN{Benjamin Naoto Chiche}
\IEEEauthorblockA{\textit{AIM, CEA, CNRS, Université Paris-Saclay} \\
\textit{Université de Paris}\\
F-91191 Gif-sur-Yvette, France \\
\textit{Safran Electronics \& Defense}\\
Massy-Palaiseau, France \\
benjamin.naoto.chiche@outlook.com}
\and
\IEEEauthorblockN{Joana Frontera-Pons}
\IEEEauthorblockA{\textit{DR2I, Institut Polytechnique des Sciences Avanceés} \\F-94200 Ivry-sur-Seine, France  \\
\textit{AIM, CEA, CNRS, Université Paris-Saclay} \\ \textit{Université de Paris}\\
F-91191 Gif-sur-Yvette, France \\
joana.frontera-pons@cea.fr}
\and
\IEEEauthorblockN{Arnaud Woiselle}
\IEEEauthorblockA{\textit{Safran Electronics \& Defense}\\
Massy-Palaiseau, France \\
arnaud.woiselle@safrangroup.com}
\and
\IEEEauthorblockN{Jean-Luc Starck}
\IEEEauthorblockA{\textit{AIM, CEA, CNRS, Université Paris-Saclay} \\
\textit{Université de Paris}\\
F-91191 Gif-sur-Yvette, France \\
https://orcid.org/0000-0003-2177-7794}
}
\IEEEoverridecommandlockouts
\IEEEpubid{\makebox[\columnwidth]{978-1-7281-8750-1/20/\$31.00~\copyright2020 IEEE \hfill} \hspace{\columnsep}\makebox[\columnwidth]{ }}
\maketitle
\IEEEpubidadjcol

\begin{abstract}
Video super-resolution (VSR) aims to reconstruct a sequence of high-resolution (HR) images from their corresponding low-resolution (LR) versions. Traditionally, solving a VSR problem has been based on iterative algorithms that can exploit prior knowledge on image formation and assumptions on the  motion. However, these classical methods struggle at incorporating complex statistics from natural images. Furthermore, VSR has recently benefited from the improvement brought by deep learning (DL) algorithms. These techniques can efficiently learn spatial patterns from large collections of images. Yet, they fail to incorporate some knowledge about the image formation model, which limits their flexibility.\\
Unrolled optimization algorithms, developed for inverse problems resolution, allow to include prior information into deep learning architectures. They have been used mainly for single image restoration tasks. Adapting an unrolled neural network structure can bring the following benefits. First, this may increase performance of the super-resolution task. Then, this gives neural networks better interpretability. Finally, this allows flexibility in learning a single model to nonblindly deal with multiple degradations.\\
In this paper, we propose a new VSR neural network based on unrolled optimization techniques and discuss its performance.

\end{abstract}

\begin{IEEEkeywords}
Video super-resolution, single-image super-resolution, unrolled optimization algorithm
\end{IEEEkeywords}

\section{Introduction}

Video super-resolution(VSR) is an inverse problem that extends single-image super-resolution (SISR). While SISR aims to generate a high-resolution (HR) image from its low-resolution (LR) version, in VSR the goal is to reconstruct a sequence of HR images from the sequence of their LR counterpart. The idea behind VSR, which makes it fundamentally different from SISR, is that the fusion of several LR images produces a HR image. Both problems were traditionally solved, as other inverse problems, using iterative algorithms based on formal optimization. These techniques exploit prior knowledge on the image formation model (and assumptions on motions for VSR~\cite{farsiu2004fast,farsiu2005multiframe,farsiu2006video,liu2011bayesian}).
Recently, both SISR and VSR have greatly benefited from deep learning (DL) methods that can efficiently learn complex statistics from natural images~\cite{kim2016accurate,shi2016real,zhang2018learning,sajjadi2018frame,jo2018deep,wang2019edvr,yi2019progressive,fuoli2019efficient}.

\noindent While classical algorithms succeed at exploiting prior information from the model, they are not suitable to derive meaningful statistics from the images and rely instead on hand-crafted priors. Opposed to this, DL techniques based on convolutional neural networks (CNNs) can learn non-trivial features without incorporating prior knowledge of the image formation model. Unrolled optimization techniques consist of unfolding the iterative loop of a classical iterative algorithm with a given number of iteration and representing all operations as layers of a neural network. This network can then be trained and optimized as any other network, learning from data while keeping the knowledge of the inverse problem in its internal structure. This technique has been successful in various inverse problems~\cite{gregor2010learning,diamond2017unrolled,adler2018learned,aggarwal2018modl,gilton2019neumann,zhang2020deep}. Unrolled networks may present the following advantages over purely learning-based networks. First, they may perform better based on certain metrics. Second, they present better interpretability. Third, they present scalability in training a single model to deal with multiple degradations in a nonblind manner, similarly to the network that was presented in~\cite{zhang2018learning}. This is all the more interesting that training a competitive VSR network is time-consuming~\cite{yi2019progressive}.

\noindent Previous works on unrolled optimization algorithms focus on single image restoration, such as image denoising, deblurring and SISR. To the best of our knowledge, unrolled optimization algorithm has never been explored to tackle VSR. On the other hand, most studies on VSR currently search for the best performing purely learning-based network under single degradation, without incorporating image formation model in the network nor dealing with multiple degradation. In this context, we introduce a framework coined Unrolled Video Super-Resolution (UVSR) that resembles FRVSR~\cite{sajjadi2018frame} but is based on unrolled optimization. In other words, we first model an image sequence formation model. Then, UVSR involves an unrolled network - more precisely unrolled gradient descent network - that is designed to solve the modeled problem. We empirically assess the UVSR performance through implementation, training on the MM522 dataset~\cite{8579237,yi2019progressive} and testing on the Vid-4 dataset~\cite{liu2011bayesian}. PSNR and SSIM are used for performance evaluation. Two experimental configurations are set up: one that involves a single degradation and another that involves multiple degradations. Both of them are noise-free. In the first one, UVSR is compared to the resembling FRVSR and three state-of-the-art (SOTA) networks. In the second one, UVSR is compared to FRVSR-MD, an improved version of FRVSR that we design so that the network can deal with multiple degradations. Qualitative evaluation is also done in the single degradation configuration. 




\section{Related work} \label{rw}

\subsubsection{Single-image super-resolution}

Solving SISR problem was traditionally based on formal optimization and iterative algorithms, but recent methods are mostly based on DL. The authors in~\cite{kim2016accurate} proposed a very deep CNN architecture with a skip connection that works in the HR image space. Alternatively, a faster CNN that mainly operates in LR image space was presented in~\cite{shi2016real}. Differently from these purely learning-based approaches, the work in~\cite{zhang2018learning} proposed to nonblindly deal with multiple degradations by concatenating to the input degraded image the degradation maps that encode the blur kernels and noise levels. Notably, they suggested to project different blur kernels by the PCA (Principal Component Analysis) technique.


\subsubsection{Video super-resolution}
The goal in VSR is to reconstruct a sequence of HR images from their LR versions. This requires realistic spatial details and temporal coherence. Classical solving of VSR problem was based on iterative algorithms that exploit knowledge and assumptions on image sequence formation model and motions~\cite{farsiu2004fast,farsiu2005multiframe,farsiu2006video,liu2011bayesian}. VSR has recently benefited from DL methods~\cite{sajjadi2018frame,wang2019edvr,jo2018deep,tian2018tdan,yi2019progressive,fuoli2019efficient}. Authors in~\cite{wang2019edvr,jo2018deep,tian2018tdan,yi2019progressive} took a batch of multiple low-resolution frames as inputs to fuse them to reconstruct
a high-resolution frame. In contrast, authors in~\cite{sajjadi2018frame,fuoli2019efficient} prioritized recursive approaches. Approaches of~\cite{sajjadi2018frame} used the previously estimated HR frame to super-resolve the subsequent LR frame. The stateful network in~\cite{fuoli2019efficient} extends this by designing a recurrent cell.

The network we design in this study is similar to FRVSR~\cite{sajjadi2018frame}. The latter involves two CNNs that work in LR space: FNet and SRNet. FNet estimates the optical flow map between the current LR frame and the previous LR frame. The flow map is then interpolated and used to warp the previously estimated HR frame. The warped HR frame is mapped to LR space based on space-to-depth transformation and fed to SRNet along with the current LR frame to operate super-resolution. The whole network is trained end-to-end based on a loss on the super-resolution task as well as an additional loss on the warped previous LR frame. 

In contrast with approaches that batch-process LR frames to
estimate a  HR frame, in FRVSR each input frame is processed only once and only
one image is warped at each step. This makes FRVSR more interesting from an application point of view which requires decreased computational cost and faster inference. Moreover, contrary to FRVSR, the batch-processing VSR methods generate independent output HR frames, which reduces temporal consistency of the produced HR frames, resulting in flickering artifacts. Finally, some approaches that present outstanding performance, for example EDVR~\cite{wang2019edvr}, are undeployable in practice because of their massive network size. FRVSR, unlike these methods, is still usable in an application with power and latency constraints.

\subsubsection{Unrolled optimization algorithm} \label{unroll}
Regarding inverse problems, classical algorithms can exploit knowledge of the image formation model while they struggle to incorporate complex statistics from natural images. Opposed to this, deep neural networks can learn spatial patterns from images without incorporating prior knowledge of the image formation model. One method that can benefit from both of these approaches is unrolled optimization algorithm. This consists of unrolling a given number of iterations of an optimization algorithm and representing all operations as layers of a neural network. This technique has been used in various signal processing problems, including inverse problems. Authors in~\cite{gregor2010learning} unrolled FISTA and ISTA for sparse coding; The work in~\cite{diamond2017unrolled} unrolled algorithms such as proximal gradient, gradient descent, ADMM and LADMM for inverse problems such as denoising or deblurring; In~\cite{adler2018learned}, primal-dual algorithms were unfolded for tomographic reconstruction. Authors in~\cite{aggarwal2018modl} unrolled an alternating minimization algorithm for MRI.

More specifically, authors in~\cite{diamond2017unrolled} introduced ODP (Optimization With Deep Priors) framework, that factorizes networks into prior or CNN step and data step. The first one encodes statistical image priors. The second one incorporates prior knowledge about the formation model and increases data fidelity. For example, in an ODP gradient descent network, the CNN represents the gradient of the prior term. Our work relies on the ODP gradient descent network. An example of an ODP gradient descent network that is designed to solve SISR problem is described in Algorithm~\ref{algo: sisr}. This network unrolls the classical gradient descent algorithm. We choose to use ODP gradient descent network rather than other ODP networks that unroll other algorithms because of its simplicity (it only involves the forward operators in the image formation model and their adjoint operators in the data step) and its fast inference speed (under the condition that the operators are fast, for example a blur by a separable kernel).

\begin{algorithm}

Input: $y, H, s$ 

Initialization; $ x ^0 = H^T B U_s y $

\For{$k\gets0$ \KwTo $K-1$}{
 
    $ z^k  \leftarrow N_{\theta_k} (x ^k) $
    
     $ x ^{k+1} \leftarrow x ^k + z^k - \alpha_k H^T B U_s D_s H x ^k + \alpha_k x ^0 $ 
     }

 Output: $ x^{K} $
  \caption{ODP gradient descent SISR network. $y$ is the observed LR image to be restored. Other notations are explained in section~\ref{method}.}
  \label{algo: sisr}
 \end{algorithm}

\section{Method} \label{method}
This section describes the proposed UVSR framework.
\subsection{Variational Methods}

We first model the VSR problem. Similarly to~\cite{farsiu2004fast,liu2011bayesian,pan2020deep}, we rely on a forward video formation model that is represented by the following equation:

\begin{equation}
y_j = D_s H F_{u_{t \rightarrow j}} x_{t} + n_j \qquad t - N \leq j \leq t + N
\label{eq: forward}
\end{equation}

where $y_j$ represents the observed $j$-th LR video frame and $x_{t}$ represents an HR reference video frame. $y_j$ is obtained by passing $x_{t}$ to the warping operator $F_{u_{t \rightarrow j}}$ with regard to optical flow $u_{t \rightarrow j}$ from $t$ to $j$, the blur operator $H$ and the downsampling operator $D_s$. $s$ is the scale factor of the downsampling. We suppose that with $s=4$, the operator $D_s = D_4$ samples every 4-th pixel in each dimension and the operator $H$ is a Gaussian blur with standard deviation $\sigma$. For the adjoint operator of $D_s H$, we use the operator $ H^T B U_s $ where $U_s$  upsamples the input by the factor $s$ by inserting zeros and $B$ replaces these zeros by bilinear interpolation. $H$ is a Gaussian blur so $H = H^T$. The adjoint operator of $F_{u_{t \rightarrow j}}$ is $F^T _{u_{t \rightarrow j}} = F _{u_{j \rightarrow t}} $. The term $n_j$ accounts for the presence of the noise, that is assumed to be additive, zero-mean and white Gaussian. Realizations of this noise are assumed to be independent and identically distributed. Assuming that the images are obtained from the same camera, operators $D_s$ and $ H $ are constant over time hence do not depend on $j$. We
also assume that the operator $H$ is space-invariant~\cite{farsiu2006video}.

Supposing $H$ is known, \eqref{eq: forward} is solved in variational methods by alternatively minimizing: 

\begin{equation}
\begin{split}
\hat{x}_{t} & = \operatorname*{arg\,min}_{x_{t}}  ||D_s H x_{t} - y_{t}|| \\
& + \sum_{j = t - N, j \neq t}^{t + N} ||D_s H F_{u_{t \rightarrow j}} x_{t} - y_{j}|| \\& + \rho(x_{t})
\label{eq: variational frame estimation}
\end{split}
\end{equation}

\begin{equation}
\hat{u}_{t \rightarrow j} = \operatorname*{arg\,min}_{u_{t \rightarrow j}} ||D_s H F_{u_{t \rightarrow j}} x_{t} - y_{j}|| + \varphi(u_{t \rightarrow j})
\label{eq: variational motion estimation}
\end{equation}

where $\rho(x_{t})$ and $\varphi(u_{t \rightarrow j})$ are hand-crafted priors on the estimated frame and optical flow.

\subsection{UVSR Framework}

\subsubsection{Algorithm}

Taking example from FRVSR~\cite{sajjadi2018frame}, if we want to recover the $t$-th HR frame by only using the $t$-th and $(t - 1)$-th LR frames, \eqref{eq: variational frame estimation} and \eqref{eq: variational motion estimation} gives:




\begin{equation}
\begin{split}
\hat{x}_t & = \operatorname*{arg\,min}_{x_t} ||D_s H x_t - y_t|| \\& + ||D_s H F_{u_{t \rightarrow t - 1}} x_t - y_{t - 1}|| \\& + \rho(x_t)
\label{eq: variational frame estimation two frames t}
\end{split}
\end{equation}

\begin{equation}
\begin{split}
\hat{u}_{t \rightarrow t - 1} & = \operatorname*{arg\,min}_{u_{t \rightarrow t - 1}} ||D_s H F_{u_{t \rightarrow t - 1}} x_{t} - y_{t - 1}|| \\& + \varphi(u_{t \rightarrow t - 1})
\label{eq: variational motion estimation two frames t}
\end{split}
\end{equation}

\begin{figure*}
\centering
\includegraphics[height=4.5cm]{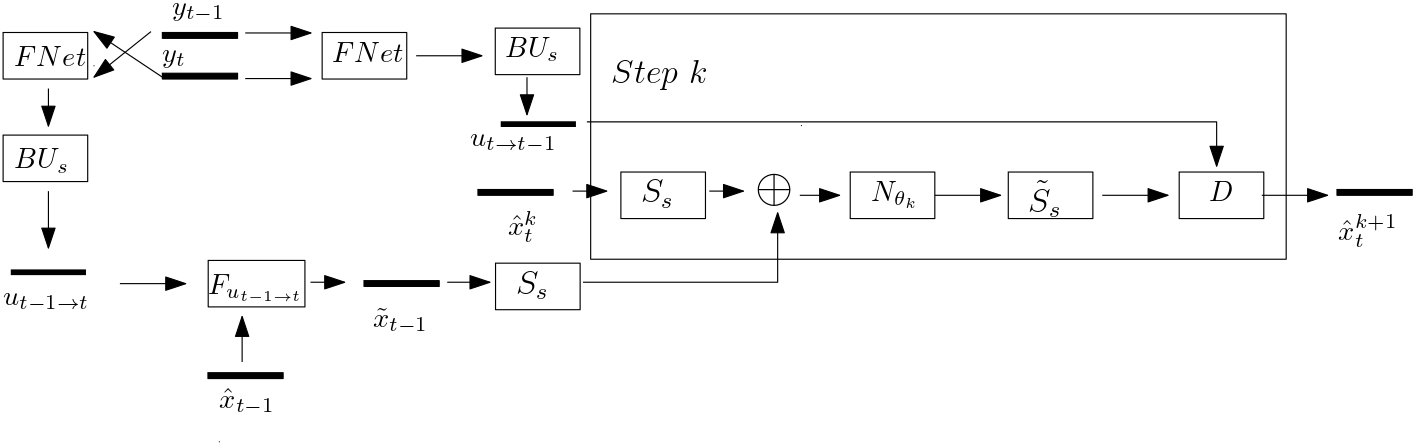}
\caption{Illustration of UVSR. $ \oplus $ denotes the concatenation in the channel dimension. D denotes the data step.}
\label{fig:UVSR blocks sd}
\end{figure*}

Therefore, to solve the VSR problem, we propose Algorithm~\ref{algo: UVSR with the motion consistency step}. Fig.~\ref{fig:UVSR blocks sd} illustrates it. $K$ is the number of unrolled iteration blocks and $\alpha_k$ and $\beta_k$ are trainable stepsize parameters of unrolled networks that are specific to each iteration block. We initialize them for the $k$-th iteration block with the following scheme: $ ( \alpha_k, \beta_k ) = ( \frac{1}{2^k}, \frac{1}{2^k} )$. $S_s$ and $\tilde{S}_s$ are respectively space-to-depth and depth-to-space~\cite{sajjadi2018frame,shi2016real} transformations. They help decreasing computational cost by allowing convolution operations to be done in the LR image space. The first operation is a mapping from HR space to LR space. The second one is its inverse. The notation $(\cdot, \cdot)$ denotes the concatenation operation.  

The detailed workflow of the algorithm is the following: the initial approximation of super-resolved frame is obtained by backprojection (line~\ref{lst:line:init}). Besides, two inferences of FNet gives $ u_{t \rightarrow t - 1, LR} $ and $ u_{t-1 \rightarrow t, LR} $, the LR optical flow maps between the two consecutive LR frames $y_{t-1}$ and $y_{t}$. These maps are then upsampled via bilinear interpolation which gives HR flow maps $ u_{t \rightarrow t - 1}$ and $ u_{t-1 \rightarrow t}$. The HR estimated of the previous frame ${x}_{t-1}$ is then warped according to $ u_{t-1 \rightarrow t}$ which produces $ \tilde{x}_{t-1}$. The latter is then space-to-depth transformed which gives $ \tilde{x}_{t-1, LR}$.

The next part of the algorithm constitutes the unrolling part. Each iteration step $k$ outputs an estimation of the super-resolved current frame $\hat{x}_t ^{k+1}$. This first involves the prior step in which the CNN $N_{\theta_k} (.)$ in same time operates fusion and encodes statistical image prior. The network takes as input the HR image estimated in the previous iteration step that is space-to-depth transformed $\hat{x}_{t,LR} ^k$ and $ \tilde{x}_{t-1, LR}$. Indeed, the prior image is drawn from a distribution with parameters $\theta$ conditioned by $\hat{x}_{t-1}$, which enforces temporal coherence. After this prior step comes the data step (line~\ref{data2}). Solving \eqref{eq: variational frame estimation two frames t} with the gradient descent unrolling algorithm involves two data consistency terms : one that is related to the current frame at $t$ ($\alpha_k H^T B U_s ( D_s H \hat{x}_t ^k - y_t)$) and another term that is related to the previous frame at $t-1$ ($\beta_k F_{u_{t -1 \rightarrow t}} H^T B U_s (D_s H F_{u_{t \rightarrow t - 1}} \hat{x}_t ^k - y_{t-1})$).

As in the data step $z_k$ is reused, the network involves a (partial) residual connection, which facilitates the flow of gradient. The latter is "partial" in the sense that the term $\hat{x}_{t-1}$ in the input that is concatenated is not involved in this connection. Space-to-depth and depth-to-space transformations are simple pixel rearrangement operations, so this connection can be kept.

One can notice the similarity of this algorithm with FRVSR~\cite{sajjadi2018frame}.

\begin{algorithm}

Input: $y_t, y_{t-1}, H, s$ 

Initialization; $  \hat{x}_t ^0 = H^T B U_s y_t $ \label{lst:line:init}

$ u_{t \rightarrow t - 1, LR}  \leftarrow  FNet(y_{t-1}, y_t) $ \label{lst:line:A1tt1}

$ u_{t -1 \rightarrow t, LR}  \leftarrow  FNet(y_t, y_{t-1}) $ \label{lst:line:A1t1t}

$ u_{t \rightarrow t - 1} \leftarrow B U_s u_{t \rightarrow t - 1, LR} $

$ u_{t-1 \rightarrow t} \leftarrow B U_s u_{t -1 \rightarrow t, LR} $

$ \tilde{x}_{t-1} \leftarrow F_{u_{t - 1 \rightarrow t}} \hat{x}_{t-1} $

$ \tilde{x}_{t-1, LR} \leftarrow S_s (\tilde{x}_{t-1}) $

 \For{$k\gets0$ \KwTo $K-1$}{
		
	$ \hat{x}_{t, LR} ^k \leftarrow S_s (\hat{x}_t ^k) $  
 
    $ z^k _{LR} \leftarrow N_{\theta_k} (\hat{x}_{t, LR} ^k, \tilde{x}_{t-1, LR}) $
	
	$ z^k \leftarrow \tilde{S}_s (z^k _{LR}) $    
    
    $ \hat{x}_t ^{k+1} \leftarrow \hat{x}_t ^k + z^k - \alpha_k H^T B U_s ( D_s H \hat{x}_t ^k - y_t) - \beta_k F_{u_{t -1 \rightarrow t}} H^T B U_s (D_s H F_{u_{t \rightarrow t - 1}} \hat{x}_t ^k - y_{t-1})$ \label{data2}
    }
    
$ \hat{x}_t \leftarrow \hat{x}_t ^{K} $ 

Output: $ \hat{x}_t,  u_{t -1 \rightarrow t, LR}, u_{t \rightarrow t - 1, LR}$
 \caption{UVSR}
 \label{algo: UVSR with the motion consistency step}
\end{algorithm}

\subsubsection{Architecture of the CNN} \label{dk}

Considering Algorithm~\ref{algo: UVSR with the motion consistency step}, for each of the iteration steps $k \in 0, ..., K-1$, with K being the number of total unrolled steps, the same architecture of $N_{\theta_k}(., .)$ is used. We use an architecture similar to the VDSR~\cite{kim2016accurate}, except that here, in accordance with Algorithm~\ref{algo: UVSR with the motion consistency step}, the input and output are in LR image space and there is the aforementioned partial skip connection. The choice of the two hyperparameters $ d $ and $ K $ and the number of filters in each convolutional layers $f$  has a significant impact on the inference speed and number of parameters of UVSR. Our proposed method is similar to the FRVSR~\cite{sajjadi2018frame} that we consider as a baseline. For fair comparison and to demonstrate the benefit brought from the unrolled architecture, we choose $d$, $K$ and $f$ so that the number of parameters of UVSR becomes a bit smaller than the number of parameters of FRVSR. Therefore, we choose $d = 7$, $ K = 3 $ and $ f=128 $ . Table~\ref{table:eval} shows numbers of parameters of FRVSR and UVSR.

\subsubsection{Architecture of FNet}

We use the same architecture of FNet as in~\cite{sajjadi2018frame}. This contributes to a fair comparison between FRVSR and UVSR. 


\subsubsection{Loss functions}

We use the following loss function:  

\begin{equation}
\begin{split}
\mathcal{L} & = \mathcal{L}_{sr} + \mathcal{L}_{flow_{t-1 \rightarrow t}} +\mathcal{L}_{flow_{t \rightarrow t-1}} \\& 
 = ||\hat{x}_{t} - x_t||_2 ^2 + ||F_{u_{t-1 \rightarrow t, LR}} y_{t-1} - y_t||_2 ^2 \\ 
& + ||F_{u_{t \rightarrow t - 1, LR}} y_{t} - y_{t - 1}||_2 ^2 
\end{split}
\label{loss1}
\end{equation}

The first term is related to the super resolution task. The second and third terms account for the optical flow estimation from FNet. All losses are backpropagated through both FNet and the unrolled network as well as through time and UVSR is end-to-end trained, as in~\cite{sajjadi2018frame}.

\section{Experimental results} \label{section: Experiment}

This section describes how we assess the UVSR performance via experiment.

\subsection{Datasets}

We use the MM522 dataset~\cite{8579237,yi2019progressive} for training and the Vid-4 dataset~\cite{liu2011bayesian} for testing. In the training phase, clips of 10 frames are randomly cropped with crop size $256 \cdot 256$ from the dataset. To generate LR frames, each HR video frame in the datasets is firstly blurred by a Gaussian kernel of standard deviation $\sigma$, then downsampled by sampling every 4-th pixel in each dimension (the sampling factor being $s = 4$), which generates LR video frames. No noise is added after. The variance of the Gaussian noise $n_t$ in \eqref{eq: forward} is therefore zero. We set up two experimental configurations. In the first configuration, $\sigma $ is fixed to $1.6$ for both train and test sets, which enables to adopt similar experimental condition as in~\cite{jo2018deep,yi2019progressive,sajjadi2018frame}. We deal with a single degradation in this case. In the second configuration, for each sequence $\sigma = \sigma_{train}  $ is uniformly sampled between $ 0.375 $ and $ 2.825 $ in the training phase. The value $\sigma = \sigma_{test} =  1.6$ is chosen for the test set. This allows to assess a single UVSR under multiple degradations. Data augmentation by random flipping is operated during training.


\subsection{Evaluation}

\subsubsection{Performance metrics}

We use similar evaluation protocol as in~\cite{7444187,jo2018deep,yi2019progressive} over the test set. PSNR and SSIM are computed over video pixels on the brightness channel of the ITU-R BT.601 YCbCr standard, excluding first and last two frames and border pixels (8 pixels). We also compute number of parameters and measure testing time cost as being the time needed to generate one $ 1920 \cdot 1080 $ frame when upscaling factor is 4. We perform experiments with an Intel I7-8700K CPU and one NVIDIA GTX 1080Ti GPU.

We compare our UVSR in the first single degradation configuration to FRVSR and the SOTA networks (DUF~\cite{jo2018deep}, PFNL~\cite{yi2019progressive}, and RLSP~\cite{fuoli2019efficient}). In this study we do not consider EDVR~\cite{wang2019edvr} because of its too large size. For DUF and PFNL, we use values that are reported in~\cite{yi2019progressive} as performances of these networks were measured based on the same setting. 
We implemented FRVSR and RLSP to measure their performances as the configurations in~\cite{sajjadi2018frame,fuoli2019efficient} are different from ours.

For the second, multiple degradations configuration, we compare UVSR with an FRVSR that we improve in order to account for the multiple degradation configurations. We coin it FRVSR-MD. The latter is derived from FRVSR but the stretched feature maps that encodes knowledge about $\sigma$ are also concatenated at the input of SRNet, similarly to~\cite{zhang2018learning}. PCA is used to reduce the dimensionality of blur kernels to 10. 

\subsubsection{Qualitative evaluation}

We also qualitatively evaluate UVSR based on generated predictions and temporal profiles.

\subsection{Results}

Table~\ref{table:eval} summarizes comparisons between UVSR and other networks under single degradation. We observe that in terms of PSNR, on average UVSR does not significantly improve over FRVSR and performs worse than the SOTA networks. However, for the~\textit{foliage} sequence, UVSR performs the best. Regarding SSIM, on average UVSR performs better than DUF and FRVSR and worse than RLSP and PFNL. Here also, for the~\textit{foliage} sequence UVSR performs the best. We note that the PSNR and SSIM values of FRVSR that we implemented are higher than the ones that are reported in~\cite{yi2019progressive} that adopted similar experimental configuration. Thus we make a fair comparison. Then, UVSR presents faster inference than FRVSR and the SOTA networks except RLSP. Regarding model complexity, UVSR presents less parameters than FRVSR and DUF but more parameters than PFNL and RLSP.

Table~\ref{table:evalmd} compares UVSR and FRVSR-MD under multiple degradations. One can see that UVSR has less parameters than and is faster than FRVSR-MD. Concerning PSNR, UVSR does not improve over FRVSR-MD. But UVSR outperforms FRVSR-MD with respect to SSIM.


\setlength{\tabcolsep}{4pt}
\begin{table}
\begin{center}
\caption{Number of parameters, testing time and PSNR(dB)/SSIM of different models on Vid4 test set with $\sigma = 1.6$. Values related to networks with '*' are taken from the referred publication. We implemented networks without '*'. Bold indicates the best performance.}
\label{table:eval}
\scalebox{0.8}{\begin{tabular}{llllll}
\hline\noalign{\smallskip}
Sequences & FRVSR & UVSR & *DUF~\cite{yi2019progressive}& *PFNL~\cite{yi2019progressive} & RLSP~\cite{fuoli2019efficient}\\
\noalign{\smallskip}
\hline
\noalign{\smallskip}
Calendar & 23.90/0.8092 & 24.03/0.8102 & 23.85/0.8052 & \textbf{24.37}/\textbf{0.8246} & 24.08/0.8160 \\
City  & 27.79/0.8220 & 27.59/0.8209 & 27.97/0.8253 & 28.09/\textbf{0.8385} & \textbf{28.10}/0.8278 \\
Foliage  & 26.53/0.7806 & \textbf{26.58}/\textbf{0.7819} & 26.22/0.7646 & 26.51/0.7768 & 26.43/0.7766 \\
Walk  & 29.98/0.9029 & 30.03/0.9035 & 30.47/0.9118 & \textbf{30.65}/\textbf{0.9135} & 30.36/0.9101 \\
Average  & 27.05/0.8287 & 27.06/0.8291 & 27.13/0.8267 & \textbf{27.40}/\textbf{0.8384} & 27.24/0.8326 \\
\hline\noalign{\smallskip}
\# param. (M) & 5.055 & 4.624 &  5.829 & \textbf{3.003} & 4.225 \\
\hline\noalign{\smallskip}
Testing time (ms) & 214 & 170 &  2754 & 741 & \textbf{80} \\
\end{tabular}}
\end{center}
\end{table}
\setlength{\tabcolsep}{1.4pt}

\setlength{\tabcolsep}{4pt}
\begin{table}
\begin{center}
\caption{Number of parameters, testing time and PSNR(dB)/SSIM of FRVSR-MD and UVSR on Vid4 test set with $\sigma_{train} \in [0.375, 2.825]$ and $ \sigma_{test} = 1.6 $. Bold indicates the best performance.}
\label{table:evalmd}
\begin{tabular}{lll}
\hline\noalign{\smallskip} 
 Sequences & FRVSR-MD & UVSR\\
\noalign{\smallskip}
\hline
\noalign{\smallskip}
Calendar & 23.52/0.7870 & \textbf{23.62}/\textbf{0.7947}  \\
City & \textbf{27.64}/0.8093 & 27.60/\textbf{0.8099}  \\
Foliage & \textbf{26.17}/0.7596 & 26.05/\textbf{0.7614}  \\
Walk & 29.68/0.8971 & \textbf{29.74}/\textbf{0.8982} \\
Average & \textbf{26.75}/0.8132 & \textbf{26.75}/\textbf{0.8161} \\
\hline
\# param. (M) &  5.066 & \textbf{4.624} \\
\hline
Testing time (ms) &  216 & \textbf{170} \\
\end{tabular}
\end{center}
\end{table}
\setlength{\tabcolsep}{1.4pt}

From the above observations, the situation UVSR is the most adapted seems to be in the presence of multiple degradations with constraints on model size and inference latency. Moreover, UVSR rather presents more satisfactory performance with respect to SSIM than PSNR. This being said, SSIM is a better measure of perceived visual quality than PSNR~\cite{wang2004image}. The following observations are in the same vein.

\begin{figure} 
    \centering
  \subfloat[GT\label{1a}]{%
       \includegraphics[width=0.45\linewidth, trim={7cm 4cm 8cm 10cm}, clip]{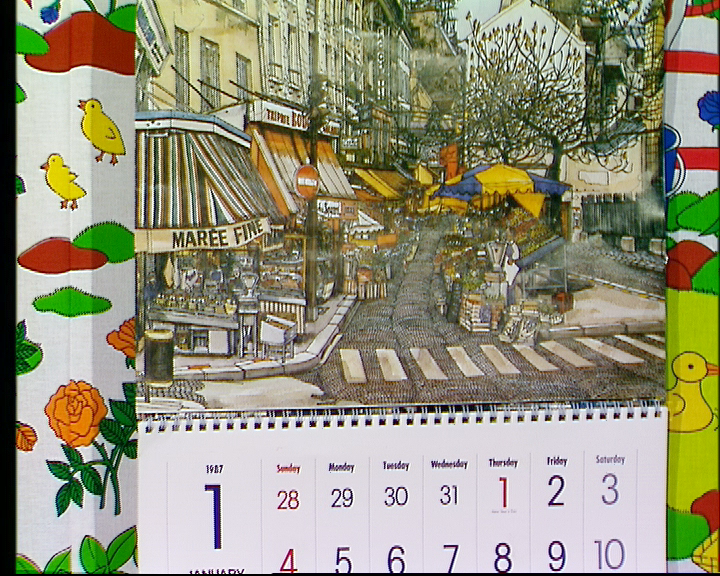}}
    \hfill
  \subfloat[FRVSR\label{1b}]{%
        \includegraphics[width=0.45\linewidth, trim={7cm 4cm 8cm 10cm}, clip]{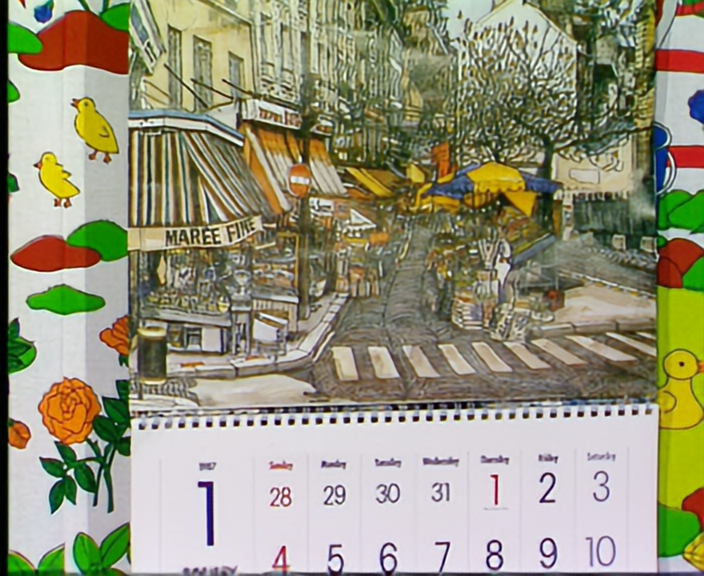}}
    \\
  \subfloat[UVSR\label{1c}]{%
        \includegraphics[width=0.45\linewidth, trim={7cm 4cm 8cm 10cm}, clip]{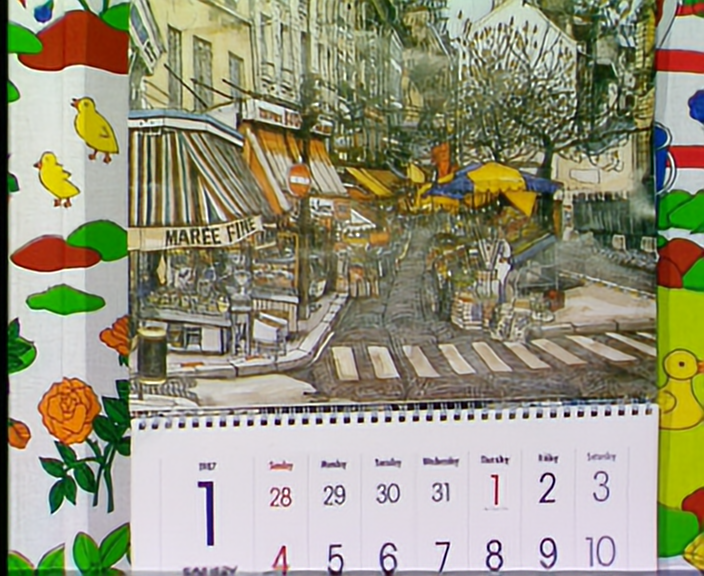}}
  
  \caption{Visual comparison on the \textit{calendar} sequence from Vid4.}
  \label{fig:visualCalendar} 
\end{figure}

Fig.~\ref{fig:visualCalendar} visually compares UVSR and FRVSR under single degradation. One can see that for the frame from the~\textit{calendar} sequence, UVSR more sharply estimates the clips and the road.

\begin{figure} 
    \centering
    (a)\includegraphics[width=0.95\linewidth, trim={0cm 0cm 11cm 0cm}, clip]{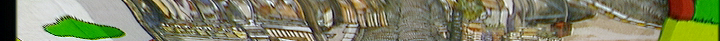}
    \\
    (b)\includegraphics[width=0.95\linewidth, trim={0cm 0cm 11cm 0cm}, clip]{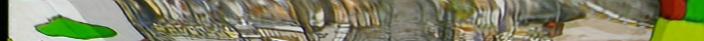}
    \\
    (c)\includegraphics[width=0.95\linewidth, trim={0cm 0cm 11cm 0cm}, clip]{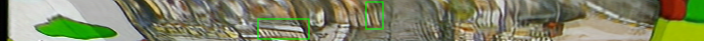}
    \caption{Temporal profiles on the \textit{Calendar} sequence. (a-c) are respectively GT, FRVSR and UVSR.}
    \label{fig:temporal_profile}
\end{figure}

Fig.~\ref{fig:temporal_profile} shows temporal profiles for this~\textit{calendar} sequence. We observe that UVSR produces sharper results than FRVSR. This shows that UVSR enables better temporal coherence and we think this is due to the fusion and the presence of a motion related data consistency term at each unrolled iteration step.

%

\section{Conclusions} \label{conclusions}

We introduced UVSR, a DL-based VSR framework based on unrolled gradient descent algorithm. UVSR, in contrast to purely DL-based VSR methods, can incorporate prior knowledge about image degradation model and enables a better interpretability of the role of CNNs based on its correspondence to classical iterative algorithms. We compared UVSR with FRVSR, three SOTA networks and FRVSR-MD under single/multiple degradation configurations, considering PSNR and SSIM over the test set, number of parameters, inference speed, visual evaluation and temporal coherence. We find that:

\begin{itemize}
    \item UVSR is faster than FRVSR, FRVSR-MD and the SOTA networks except one;
    \item UVSR presents less parameters than FRVSR, FRVSR-MD and a SOTA network but more parameters than the other two SOTA networks;
    \item Based on average PSNR, UVSR does not improve over FRVSR and performs worse than the SOTA networks under single degradation. UVSR does not improve over FRVSR-MD under multiple degradations;
    \item Regarding average SSIM, UVSR performs better than FRVSR and a SOTA network and performs worse than the two other SOTA networks when a single degradation is involved. When there are multiple degradations, UVSR improves over FRVSR-MD;
    \item For a test sequence, UVSR performs the best in terms of both PSNR and SSIM under single degradation;
    \item UVSR sharply restores images and shows high temporal coherence. 
\end{itemize}


Empirical evaluation seems to indicate the situation UVSR is the most adapted is when there are multiple degradations with constraints on inference speed and number of parameters.

\section{Future work}


ODP networks based on other algorithms (e.g., proximal gradient, ADMM, ...) could be also adapted for VSR. This would introduce different advantages and disadvantages regarding enabled data fidelity as well as computational cost. In the same idea, other unrolling networks based on iterative algorithms (e.g., primal-dual algorithms~\cite{adler2018learned}), as well as other unrolling networks that present notably different unrolling schemes (e.g., Neumann networks~\cite{gilton2019neumann}) could also be used in our UVSR.


Finally, in our study FNet for motion compensation and the unrolled network were decoupled. An interesting perspective would be to investigate whether it would be possible to unify them, by designing for example a VSR unrolling network alone that operates both motion estimation and frame alignment within its iteration blocks.

\bibliographystyle{IEEEtran}
\bibliography{IEEEabrv,MyThesis}


\end{document}